\documentclass[fleqn,10pt]{wlscirep}
\usepackage[utf8]{inputenc}
\usepackage[T1]{fontenc}
\usepackage{bm}
\usepackage{comment}
\usepackage{booktabs}  
\usepackage{multirow}  
\usepackage{graphicx}  
\usepackage{enumitem}  
\usepackage{float}

\title{Power Distribution System Blackstart Restoration Using Renewable Energy}

\author[1]{Wenlong Shi}
\author[1]{Hongyi Li}
\author[1]{Cong Bai}
\author[1,*]{Zhaoyu Wang}
\affil[1]{Iowa State University, Department
	of Electrical and Computer Engineering, Ames, IA, USA.}

\affil[*]{Corresponding Author: Zhaoyu Wang (wzy@iastate.edu).}

\begin{abstract}
	
Integrating renewable energy sources into the grid not only reduces global carbon emissions, but also facilitates distribution system (DS) blackstart restoration. This process leverages renewable energy, inverters, situational awareness and  distribution automation to initiate blackstart at the DS level, obtaining a fast response and bottom-up restoration. In this Review, we survey the latest technological advances for DS blackstart restoration using renewable energy. We first present mathematical models for distributed energy resources (DERs), network topology, and load dynamics. We then discuss how the situational awareness can help improve restoration performance through real-time monitoring and forecasting. Next, the DS  blackstart restoration problem, including objectives, constraints, and existing methodologies for decision-making are provided. Lastly, we outline remaining challenges, and highlight the opportunities and future research directions.

\end{abstract}
\begin{document}

\flushbottom
\maketitle

\thispagestyle{empty}

\section*{Introduction}

Large-scale blackouts caused by natural disasters pose substantial risks to modern power systems \cite{xu2024resilience}. Between 2000 and 2021, approximately 83\% of major blackouts in the U.S. are  attributed to extreme weather events. Also, since 2011, their annual frequency is increased by nearly 80\% \cite{ClimateCentral}. High-impact incidents, such as the 2021 Texas winter storm \cite{lee2022community}, 2022 Hurricane Ian \cite{entress2023public}, 2024 Hurricane Helene \cite{Helene} that left millions without power, have demonstrated the severity. Accordingly,  blackstart restoration has emerged as a hot topic both in industry and academic.

\begin{figure}[H]
	\centering
	\includegraphics[width=7in,trim={0cm 17.5cm 0cm 1cm},clip]{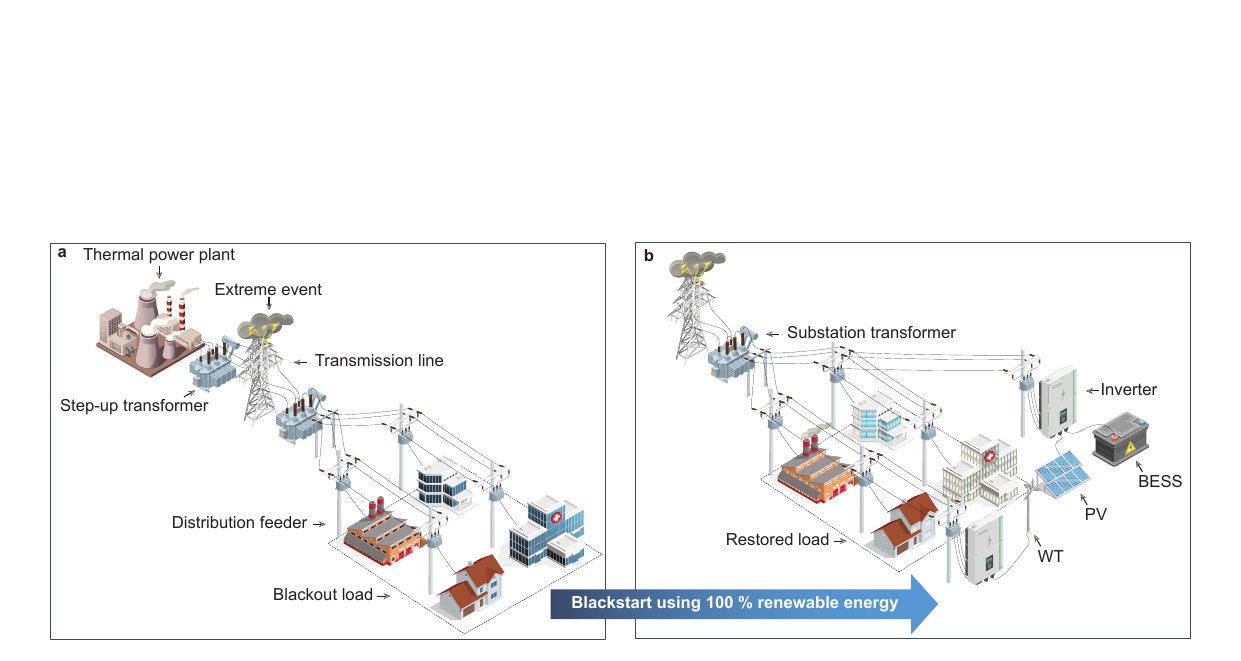}
	\caption{\textbf{Evolution of blackstart restoration.} In traditional power systems, blackstart is initiated by large thermal power plants, which supply power through a step-up transformer to the transmission system. The power flows through transmission lines to substation transformers, which then energize distribution feeders in a top-down restoration sequence. Finally, industrial, commercial, and residential loads are gradually restored. In modern power systems, blackstart restoration using distributed energy resources (DERs) follows a bottom-up approach. DERs first restore loads nearby, then expand their coverage to bring more DERs and loads back online.}
	\label{fig1}
\end{figure}

Traditional blackstart restoration employs an intuitive top-down principle. It starts the restoration from large power plants, then restores the transmission network, and at last re-energizes downstream distribution systems (DSs)  \cite{liang2024power}. However, this top-down principle faces increasing limitations as power system evolved. First, restoration relying on power plants are vulnerable to bulk system damage. The long distance power supply can be easily hindered, and the end users can not be restored. Second, this approach is not designed for power systems where distributed energy resources (DERs) gradually becomes dominated. In other words, the decentralized nature of DERs should be leveraged, especially its advantage of being closer to the end users. Third, large power plants are often powered by fossil fuels. Traditional blackstart restoration can not well utilize renewable energy, which contradicts global efforts to reduce carbon emissions (Fig. \ref{fig1}a).

According to IEEE Std 2800-2022 \cite{9762253}, DERs refer to solar photovoltaics (PVs), wind turbines, energy storage systems (ESSs), or their combinations. To meet national decarbonization targets, such as achieving 100\% clean electricity by 2035 in the U.S. and climate neutrality by 2050 in the EU \cite{USDecarbon, ENDecarbon}, DERs are increasingly integrated at the distribution level worldwide. These DERs connect to the grid through inverters, classifying them as inverter-based resources \cite{matevosyan2021future}.With advancements in power electronics, DERs can execute controlled power injection with voltage and frequency regulation. This makes researchers redefine DERs' capabilities and revisit their potentials, especially in the context of blackstart restoration. 

The blackstart restoration using DERs is also consistent with IEEE Std 1547.4-2011 \cite{5960751}, which provides guidelines for islanded system operation, and enables DERs to achieve a bottom-up blackstart process (Fig. \ref{fig1}b). By transforming the role of distributed generation from passive load-following to active blackstart initiation, the bottom-up approach reduces reliance on centralized generation and transmission infrastructure while enhancing overall system resilience  \cite{blaabjerg2017distributed}. Despite these advantages, there are also challenges exist in blackstart restoration using DERs. First, the low inertia resulted by the synchronous generator missing increases the risk of system instability. Second, the dynamic reconfiguration of the network during a blackstart requires coordination across the entire system to maintain power and load balance. Third, the sudden load restoration, particularly during cold load pickup, requires careful sequencing to mitigate excessive voltage drops and frequency deviations.

In this Review, we examine recent technological advances towards DS blackstart restoration with renewable energy. First, we present models for DERs, network topology, and load dynamics. The role of situational awareness in improving restoration through real-time monitoring and forecasting is also discussed. Second, the DS blackstart restoration problems, including objectives, constraints, and methodologies for decision-making are reviewed. Finally, we conclude by discussing challenges of blackstart restoration with renewable energy, and present future research directions.

\section*{Models for DS Blackstart Restoration}
\begin{figure}[H]
	\centering
	\includegraphics[width=7in,trim={0cm 8.5cm 0cm 0cm},clip]{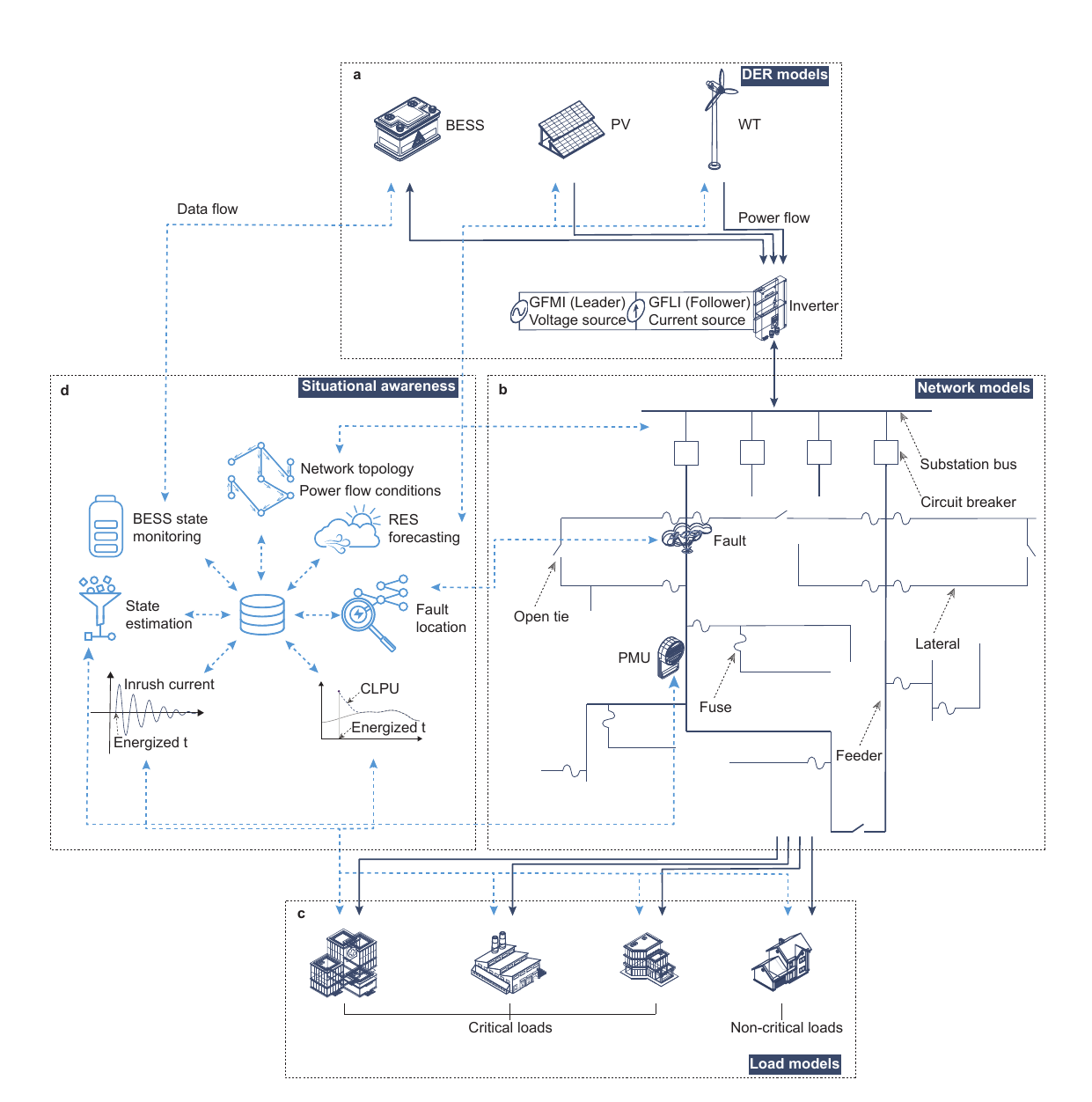}
	\caption{\textbf{Modeling framework for distribution system blackstart restoration.} a | Distributed energy resources (DERs) model: DERs participating in blackstart include, photovoltaics, and wind turbines and energy storage systems. Grid-forming inverters  operate in voltage source mode, initiating blackstart, while grid-following inverters operate in current source mode, injecting power once the network is energized. b | Network model: The network model represents the grid topology. It captures power flow dynamics, switch status, and the behavior of protection devices such as fuses and reclosers. c | Load model: Load restoration is influenced by cold load pickup, which accounts for inrush currents and increased demand after re-energization. Loads are categorized as critical and non-critical groups. d | Situational awareness: Advanced metering infrastructure and phasor measurement units provides real-time monitoring of nteworks. State estimation and fault location detection help determine the system's operational status.}
	\label{fig2}
\end{figure}

In this section, we focus on models which represent the operational characteristics of components involved in DS blackstart restoration. These models are classified into supply-side, network, and demand-side models. Also, the situational awareness in these models are discussed (Fig. \ref{fig2}).

\subsection*{Supply-Side Models}

On the supply-side, renewable energy generation and storage are two fundamental features in using solar PV, wind, and ESS. They connect to the grid through inverters which are operated in either grid-forming mode or grid-following mode.

\subsubsection*{Energy Generation and Storage Model}
Renewable energy generation refers to the process of converting natural resources into usable energy. Solar PV arrays produce direct current (DC) power from incident solar radiation, which is then converted to alternating current (AC) through an inverter. Wind turbines harness the kinetic energy of wind to drive a generator that produces AC at a specific frequency \cite{buonomano2018hybrid}. During blackstart processes, these renewable generation can be used for emergency power supply. To effectively calculate their intermittent output, effective models are developed \cite{bistline2021modeling}. These models consider the spatial variability such that geographic uncertainties are incorporated. Also, the temporal variability include both short-term and long-term are studied to model the renewable generation on daily and seasonal scales.

ESSs can facilitate an efficient utilization of renewable energy by converting electrical energy into an electrochemical form, and injecting it back into the grid when necessary\cite{akram2020review}. Discrete time-dependent modelling is often adopted to track the state-of-charge (SOC), which represents the available energy with respect to the maximum capacity \cite{ghasemi2022distribution}. In this process, energy conversion efficiencies should be included to model charging and discharging capabilities. These efficiencies coming from internal resistance and thermal effects are always considered as fixed value, even though they are related to load and battery conditions in practice \cite{wankmuller2017impact}. The charging and discharging dynamics, often expressed as C-rates, are used to represent the charging and discharging power at each time interval.

\subsubsection*{Inverter Model}

DERs connect to the grid through power electronic inverters. These inverters can be broadly classified into grid-forming and grid-following   based on their roles and control functionality (Table \ref{tab:GFMI_GFLI_Comparison}).

Grid-forming inverters (GFMIs), functioning similarly to traditional synchronous generators, serve as critical "kick-starters" to enable blackstart restoration \cite{egblack}. It can dynamically adjust power output instantaneously to deal with uncertain disturbances in load or system conditions \cite{lasseter2019grid}. In this respect, both steady-state and dynamic analyses are always considered to ensure a reliable operation of GFMIs \cite{mohammed2023grid,du2020modeling}. Specifically, steady-state analysis evaluates voltage regulation to maintain voltage levels within acceptable limits based on power sharing and synchronization. Dynamic analysis addresses how the system responds to events such as sudden load pickup, switching operations, and DER connections. Accordingly, GFMIs can be modeled as voltage sources, providing system voltage and frequency reference, and enabling grid-following inverters (GFLIs) to connect and function effectively. By maintaining a constant internal voltage phasor, GFMIs emulate the inertial response of synchronous generators, delivering instantaneous active power to counteract frequency deviations and exchanging reactive power with the grid to stabilize voltage levels \cite{rosso2021grid,meng2019fast}.

Compared with GFMIs, GFLIs function as followers during DS blackstart restoration \cite{10582520}. They can be modeled as current sources, which can control both their active and reactive power output. This is achieved by using phase-locked loops, which continuously tracks the grid's phase angle and frequency. GFLIs are operated by synchronizing with the grid voltage established by GFMIs  \cite{zarei2019control,du2020modeling,li2022revisiting}. During blackstart processes, a GFMI starts itself and establishes cranking paths to bring GFLIs back online \cite{liu2023utilizing,kuo2024building}. The decoupled control of active and reactive power ensures that once operational, GFLIs can independently manage their active and reactive power to meet grid demands flexibly \cite{liu2014decoupled}. For example, GFLIs can be operated in either grid-feeding or grid-supporting modes \cite{mandrile2020grid}. In grid-feeding mode, the maximum power point tracking (MPPT) is often employed. In grid-supporting mode, the GFLI actively regulates reactive power output to provide auxiliary services.

\subsection*{Network Models}
The network model defines the structural and electrical characteristics of DSs. It consists of two fundamental components: topology model and power flow model.

\subsubsection*{Topology Model}

The distribition network can be modeled as a graph, which consists of nodes and edges \cite{shi2024dynamic}. Specifically, nodes are used to represent components such as substations, loads, and power sources. It is characterized by attributes such as nodal voltage, load demand, and generation capacity. Except for these, electrical devices such as capacitor banks and on-load tap changers are also connected to nodes \cite{sekhavatmanesh2018analytical,macedo2021optimal}. Edges represent the physical connections between two nodes. This includes distribution lines, tie-lines, smart switches, circuit breakers, and protection devices. Voltage regulators and soft open points (SOPs), which are critical for maintaining voltage regulation, are often integrated as part of the distribution lines and tie-lines \cite{yang2022cooperative}. In addition, smart switches and circuit breakers are modeled as binary-state elements, characterized as either open or closed, directly influencing the operational and connectivity status of the network \cite{zhao2020mpc}. The parameters of edges, including impedance, power flow, capacity and thermal limits, determine the electrical performance and safe operational boundaries of the network.

\subsubsection*{Power Flow Model}

To address the nonlinearity of the standard AC power flow, two primary approaches are developed: linearization of power flow equations and convex relaxation of power flow constraints \cite{ibrahim2021low}.  Linearization foucuses on replacing the nonlinear power flow equations with linear equality constraints. Convex relaxation addresses the nonliearity by reformulating nonlinear equality constraints into convex inequality constraints. State variable is another critical consideration in power flow analysis, with two predominant models in use: the bus injection model (BIM) and the branch flow model (BFM) \cite{yang2020linear}. The BIM focuses on nodal variables such as voltages, currents, and power injections. The BFM emphasizes branch flows, such as currents and powers on individual branches.

In addition, convex relaxation approaches, such as second-order cone programming (SOCP) and semidefinite programming (SDP), are always applied. These techniques have produced various formulations, including SDP-BIM \cite{lavaei2011zero}, SOCP-BIM \cite{chen2018optimal}, SDP-BFM \cite{gan2014convex}, and SOCP-BFM \cite{farivar2013branch} with the aim of enhancing the tractability of power flow calculations. Among these, SOCP-BFM is commonly used in DS blackstart restoration problems,  due to its numerical stability and lower computational complexity \cite{gan2014exact}. Moreover, it can be extended beyond balanced systems to three-phase unbalanced systems, where distribution line segments cannot be transposed \cite{robbins2015optimal}. Another widely utilized model in DS blackstart restoration problems is the linearized DistFlow model \cite{yeh2012adaptive}, which achieves linearization by neglecting the power loss term.

\subsection*{Demand-Side Models}
Accurate load modeling is essential for managing diverse load dynamics, and accounting for cold load pickup (CLPU) is vital to prevent transient overloading that could hinder the blackstart restoration.

\subsubsection*{Load Model}
In DSs, loads can be classified according to consumer type, prioritization, switchability, and phase connection. Residential, commercial, and industrial loads differ from each other in variability and power quality requirements. For example, residential loads depend highly on user behavior and time of day, commercial loads peak during business hours, and industrial loads feature large inductive machinery  \cite{7917348}. In terms of restoration, loads are also identified as critical loads and non-critical loads. Critical loads, such as hospitals and water stations, must be restored first, while non-critical loads can be shed as needed considering energy insufficiency \cite{bie2017battling}. Moreover, switchability means that a load can be switched on and off separately. Non-switchable loads will be energized automatically when the feeder they are connected to is activated, whereas switchable loads can be picked up or curtailed flexibly \cite{zhang2020sequential}. From the phase perspective, loads are also modeled as single, two, and three-phase loads. The restoration of single-phase loads often causes phase imbalance, while three-phase loads, typical in industrial settings, often induce startup currents and require higher power quality. In addition, for load modeling, the approaches can be classified as static models and dynamic models \cite{arif2017load}. Static load model are used to represent the load demand under steady-state conditions. Dynamic load models are used to analyze time-varying load behaviors, especially during transient events. In this domain, various modeling approaches are developed including ZIP model, exponential model, frequency-dependent model, and induction motor model. Also, deep neural networks are employed recently for load prediction.

\subsubsection*{Cold Load Pickup Model}

CLPU ofen occurs after an extended outage. The reason of this phenomenon is the loss of load diversity, when multiple loads simultaneously attempt to resume operation upon re-energization \cite{xie2023dynamic}. CLPU consists of two phases: the inrush phase and the enduring phase. The inrush phase involves high inrush currents caused by transient load energization, such as motor starting and distribution transformer magnetization. This phase can last a few seconds after restoration. The enduring phase arises because of the high peaks and fluctuations caused by the reconnection of thermostatically controlled loads. This phase can last minutes to hours, with loads reaching several times normal levels at the begining of restoration. To model the CLPU, various approaches are developed. The piecewise linear model and delayed exponential model are applied to describe the CLPU curve \cite{bajic2019service,song2020robust}. However, their accuracy are limited since no uncertainties are considered. Stochastic models which incorporate randomness can provide more robust results. But they require advanced stochastic methodologies for implementation \cite{li2021restoration}.   

\subsection*{Situational Awareness}
Situational awareness in DS blackstart restoration includes real-time monitoring, predictive analysis, and adaptive control. It focuses on three  aspects: DER operations, network reconfiguration, and load behavior.

\subsubsection*{DER Operational Awareness}

Renewable energy forecasting and inverter operational monitoring are two aspects of DER operational awareness. Renewable energy forecasting leverage meteorological data, historical generation patterns, and real-time sensor measurements to improve prediction accuracy. It can be broadly categorized into statistical approaches and machine learning approaches. For statistical approaches, regression analysis and time series analysis are advantageous in linking historical data to future energy output  \cite{ahmed2019review}. However, the unexpected weather changes can deteriorate their performance. Machine learning are effective in modeling non-linear relationships in data \cite{aslam2021survey}. It can be used to  process large datasets and integrate diverse data sources, such as those incorporating spatial and temporal factors (Table \ref{tab:situational_awareness}). In addition, inverter monitoring focuses on assessing performance of both GFMIs and GFLIs \cite{qin2023integrated}. Monitoring power output and dynamic response of GFMIs can help in real-time power sharing, regulation and synchornizaiton. Monitoring the phase-locked loop of GFLIs can improve its performance in handling disturbances and avoid synchronization failures.

\subsubsection*{Network Monitoring}
Situational awareness in networks facilitates the dynamic reconfiguration of DSs during blackstart restoration. Real-time monitoring of topology changes, switching device status, and power flow conditions can help the development of restoration actions and maintain system stability \cite{chen2017modernizing}. State estimation, leveraging data from advanced metering infrastructure and phasor measurement units enhances network observability \cite{akrami2019optimal}. In addition, high-speed fault location and isolation systems can enhance the system resilience, mitigating the negative impact of outages.

\subsubsection*{Load Awareness}

Real-time monitoring and predictive modeling help accurate estimation of load recovery process and prevent supply-demand imbalances. To improve prediction accuracy, situational awareness incorporates both physics-based and data-driven modeling approaches to dynamically assess load variations \cite{li2014review} (Table \ref{tab:situational_awareness}). In physics-based models, electrical and thermal principles are used to capture the dynamics of motors, HVAC systems, and lighting. These models typically require precisely parameterizing physical systems, making them less adaptable when operating conditions change rapidly. In data-driven models, historical and real-time data are leveraged through statistical approaches and machine learning approaches to identify complex and nonlinear load behavior  \cite{zhu2022review}. These models can provide higher forecasting accuracy, yet the downside is that their performance on unseen scenarios may be limited.

\section*{Problem Formulation for DS Blackstart Restoration}

This section discusses power dispatch, voltage regulation, demand management and dynamic microgrid formation, including objective and  constraints to ensure system feasibility, stability, and operational efficiency (Table \ref{tab:problem_formulation}).

\subsection*{Problem Objectives}
The problem of DS blackstart restoration can be categorized as power dispatch, voltage regulation, demand management, and dynamic microgrid formation. The objectives can be identified as maximizing restored load, minimizing restoration time and operational costs, reducing power losses, and maintaining frequency stability.

\subsubsection*{Power Dispatch}

The challenges of power dispatch with DERs are identified as follows. First, the restoration process occurs under blackout conditions, which means there is no power delivered from bulk systems. Second, despite DERs can perform blackstart and restoration, their capability is limited. To this end, DERs, including solar PV, wind, and ESSs, must be coordinated within an integrated framework. First of all, solar and wind energy can complement each other due to their distinct generation features. For example, solar generation is highest during daylight hours, while wind generation often reaches higher levels during nighttime or in morning hours \cite{pearre2019proportioning}. This temporal complementarity can improve power supply continuously throughout a 24-hour period. Also, weather variations between solar and wind can offset each other. For example, wind generation may increase during stormy days when solar power is reduced. Secondly, renewable generation can be effectively coordinated with ESS through techniques such as peak shaving and load leveling. For example, peak shaving mitigates sharp increases in load demand by discharging ESSs during peak periods, which can reduce stress on the system \cite{di2016model}. In addition, load leveling ensures a more balanced power supply by storing excess renewable generation during off-peak periods, and discharging it during high-demand periods \cite{chen2017multi}. 

Addressing the power dispatch problem in the context of DS blackstart restoration involves balancing multiple, often conflicting objectives such as maximizing the total weighted restored load and minimizing restoration times \cite{wang2018risk,poudel2018critical}. Achieving these objectives demand the judicious allocation of limited renewable generation, and optimal scheduling switching operations to restore loads as quickly as possible \cite{wang2014coordinated,wang2015decentralized}. Also, operational constraints are important, such as ensuring thermal limits of lines are not exceeded, and maintaining voltage profiles within permissible ranges. In addition, operational costs such as ESS charging/discharging costs and renewable curtailment cost are often integrated in the objective function to achieve a cost-effective blackstart restoration \cite{carrion2023dynamic,wang2015networked}.

\subsubsection*{Voltage Regulation}

Voltage regulation during DS blackstart restoration is more complex  compared to normal conditions. First, the variability of DERs influenced by weather conditions and solar irradiance can lead to frequent voltage deviations. Second, the switch operations used for isolating faults and creating islands alter the network topology frequently. Traditional voltage regulations utilize controllable devices such as voltage regulators, capacitor banks, and on-load tap changers  \cite{sekhavatmanesh2018analytical}. Although they are helpful, their effectiveness are limited due to their discrete adjustment capabilities \cite{fan2021restoration}. In addition, the integration of ESSs in DSs offers a promising solution to voltage regulation \cite{tao2023distributed}. The capability of ESSs on flexible active and reactive power control renders dynamic respond to voltage fluctuations in real time \cite{10537984}. However, the reactive power capacity of ESS inverters can be insufficient. When ESSs are heavily loaded to meet active power demands, their ability to provide supplementary reactive power support is reduced. To maximize the voltage regulation potential of ESSs, strategies such as real power curtailment may be necessary to create additional headroom for reactive power \cite{10664453}. This trade-off highlights the need for coordinated optimization of ESS dispatch to balance active and reactive power contributions. Futhermore, SOPs represent another advanced technology for voltage regulation \cite{10521743,10508977}. When operating in voltage control mode, SOPs can deliver reactive power support without requiring active power curtailment. When operating in power flow control mode, SOPs can manage real and reactive power transfers bidirectionally between feeders, effectively balancing voltages across the network and mitigating the impacts of topology changes. The objectives of voltage regulation during DS blackstart restoration are minimizing total network power losses and minimizing voltage deviation. Minimizing total network power losses can improve power delivery efficiency while enhancing voltage profiles \cite{sekhavatmanesh2018analytical}. And, minimizing voltage deviation helps protect equipment from damage caused by over-voltage or under-voltage conditions \cite{ko2018coordinated}. It also prevents resource tripping, as DERs are highly sensitive to voltage fluctuations. 

\subsubsection*{Demand Management}

Demand response is an effective strategy of demand-side management  that directly adjusts energy consumption to align with limited supply capabilities \cite{fan2021restoration}. Two primary mechanisms are commonly used for demand response: load curtailment and load shifting. Load curtailment refers to the optimal utilization of limited power by supplying critical loads, while temporarily shedding or deferring non-essential loads \cite{jibran2021demand,hafiz2019utilising}. In contrast, load shifting involves rescheduling energy-intensive activities to periods of lower demand or higher renewable energy availability, which effectively mitigates peak load pressures  \cite{gilani2022microgrid,zhu2021co}. The demand response problem can be formulated with the objective of minimizing multi-period weighted load curtailment. This formulation prioritizes critical loads by assigning higher weights, while also accounting for the temporal dynamics of demand across a defined restoration horizon \cite{wang2021multi}. Achieving an effective redistribution of the load curve requires the incorporation of capacity constraints and load constraints such that the supply-demand balance is maintained at each period.

CLPU introduces an additional challenge in demand management. Unlike demand response which is a controlled process, CLPU results in an uncoordinated surge in load demand following blackstart restoration after a prolonged outage. The CLPU phenomenon introduces several challenges. The sudden demand spike caused by the simultaneous reactivation of previously disconnected loads can severely impact system stability and resource allocation. In addition, the magnitude and duration of CLPU are uncertain, because they depend on factors such as ambient temperature, outage duration, and load composition \cite{li2021restoration}. To address these challenges, sequential restoration is studied which aims to reconnect loads in smaller and prioritized groups rather than energize all loads simultaneously \cite{wang2023sequential,xie2023dynamic}. By staging the restoration process, the adverse effects of both the inrush and enduring phases of CLPU are mitigated, as the system is allowed to stabilize incrementally between restoration steps \cite{pang2023dynamic}.

\subsubsection*{Dynamic Microgrid Formation}

DS blackstart restoration cosidering DERs, network and load as a whole is challenging. First, the low inertia characteristics limits the system ability to handle voltage and frequency deviation. Second, inrush current induced by frequent switching operations can not be ignored. Third, disturbance events such as sudden load pickup or shedding, including those associated with CLPU, place additional stress on DERs. To overcome these issues, dynamic microgrid formation is investigated\cite{10858306,serban2017microgrid}. It aims to schedule the restoration sequence, including the start-up of DERs with GFMIs, network reconfiguration, the activation of GFLIs, load reconnection, and microgrid synchronization \cite{du2019dynamic}. Specifically, the process begins with the blackstart operation initiated by DERs with GFMIs, which establish stable voltage and frequency within a self-sustaining microgrid \cite{10858306}. Once the initial microgrid is stabilized, its electrical boundaries are dynamically expanded through network reconfiguration to establish cranking paths. This enables additional GFLIs to be brought online, and contribute to the blackstart restoration. Finally, multiple microgrids are synchronized to form larger operational islands \cite{wang2015self}. The objectives of dynamic microgrid formation involve maximizing the restored load while maintaining frequency stability and minimizing the supply-demand imbalance at each step. In addition, constraints specific to each dynamic step should be enforced, especially frequency and switching constraints must be integrated.

\subsection*{Constraints}

To ensure feasibility, stability, and reliability in DS blackstart restoration, 
constraints are identified as generation constraints, operational constraints, topology constraints and protection constraints.

\subsubsection*{Generation Constraints}
Generation constraints define the operational limits of DERs in DS blackstart restoration. Main constraints are as follows.

1) Generation availability: Weather conditions, such as sunlight and wind speed, will affect the availability of renewable energy generation \cite{kaabeche2011sizing}. For example, solar panels cannot generate power at night and produce less under overcast skies or shading. Wind turbines function within specific wind speed ranges, shutting down when speeds are below the cut-in threshold or above the cut-out limit.

2) Capacity: The capacity constraint refers to the maximum output limitations of DERs, determined by their rated capabilities \cite{samani2019tri}. For PV, wind tubine, and ESS, the capabilities are limited by the design of the panels, turbines and batteries. 

3) ESS operation: The SOC of ESS must be maintained within a permissible range to prevent overcharging or deep discharging. The charging and discharging power must be constrained by the maximum C-rate \cite{10184456}. 

4) Spinning reserve: The absence of mechanical inertia in DERs necessitates that inverters sustain reserve output over a specified duration \cite{ye2020resilient}. Effective spinning reserve implementation often requires coordination among DERs. This coordination introduces interdependent constraints on storage capacity, SOC, and C-rates.

5) Ramp rate: Ramp rate defines the maximum speed at which a power source can increase or decrease its output \cite{kang2021impact}. This constraint is critical for maintaining a proper balance between power supply and demand to minimize large transients.

6) Inverter output: The output power of inverters is constrained by their design, which specifies the maximum apparent power they can deliver \cite{xu2021review}. This limitation is necessary to keep inverters within thermal and electrical thresholds. 

\subsubsection*{Operational Constraints}

Operational constraints establish the allowable operating conditions  to maintain the stability and reliability of DSs during blackstart restoration. Main constraints are as follows.

1) Line capacity: Line capacity is the maximum power a power line can carry without exceeding its  operational limits \cite{9707495}. It is primarily influenced by factors such as conductor size, material, length, and voltage levels. Exceeding these limits can lead to equipment damage, reduced lifespan, and overall system instability.

2) Thermal limits: Thermal limits is the maximum current a line can carry without exceeding its temperature rating, which is a critical determinant of line capacity \cite{8362717}. These limits are influenced by ambient temperature and wind speed. In addition, thermal limits set the upper boundary for line capacity.

3) Nodal voltage: Voltage levels at each node should be maintained within acceptable ranges \cite{9707495}. Deviations from these limits can disrupt the operation of equipment and degrade power quality.

4) Load limits: The largest amount of load that can be picked up in a single restoration step should be limited \cite{10012033}. It is influenced by the available generation capacity and the requirement to balance power supply and demand. 

5) Frequency stability: Limiting the rate of change of frequency (RoCoF), maintaining the frequency nadir above thresholds, and achieving a quasi-steady-state frequency (QSS) within acceptable ranges \cite{10700608} is important. Especially in low-inertia systems, these limits help mitigate frequency dips caused by sudden demand surges or CLPU.

6) Synchronization: When interconnecting two microgrids, synchronization should be considered \cite{10858306}. Allowing only small deviations from the set points in terms of voltage magnitudes, frequencies, and phase angles is necessary \cite{RAI2025111434}.

\subsubsection*{Network Constraints}
Network constraints define the structural and connectivity requirements of the distribution network during blackstart restoration. Main constraints are as follows.

1) Radiality maintenance: A commonly used approach to develop radiality constraints is to leverage the parent-child node relationship. These constraints avoid loops by enforcing that every non-root node has exactly one parent, thereby preserving a radial structure \cite{chen2015resilient}. To improve computational efficiency, some concepts in graph theory in maintaining spanning trees can be used as constraints, including single-commodity flow \cite{ding2017new}, multi-commodity flow \cite{lei2020radiality}, and maximum density \cite{sun2022novel}. Furthermore, hybrid constraints that combine the parent-child node relationship and graph theory have demonstrated effectiveness, particularly in large-scale distribution networks \cite{pang2022formulation}.

2) Safety: Safety constraints govern the operation of switching devices, such as smart switches and circuit breakers, based on network conditions. Without considering synchronization, a switch can only be closed when supplying power to an unenergized area to prevent tripping issues \cite{9027100}.  Furthermore, switch operations must account for the safety of crews by ensuring no work is being conducted on the downstream network during energization \cite{chen2018toward}.

\subsubsection*{Protection Constraints}
During dynamic DS blackstart restoration, protection constraints ensure
proper coordination among protective devices such as fuses, reclosers and relays, thereby maintaining both fault sensitivity and selective operation.

1) Fuse coordination: Fuses are thermal devices whose operation is governed by their melting curves, typically defined by the minimum melting time and the maximum clearance time as functions of the fault current. To achieve selectivity between two fuses installed in series, the constraints must enforce that the backup fuse delays its operation relative to the primary fuse \cite{alam2018optimum}. 

2) Recloser coordination: Reclosers operate automatically in fast and slow tripping modes. The fast mode clears transient faults with minimal delay, while the slow mode allows time for downstream protective devices to act before the recloser trips again. To ensure selective protection, the delay between the recloser’s slow and fast operations must meet the critical time interval. This allows the fast operation to clear transient faults before the slower mode is participated as a backup.

3) Relay coordination: Overcurrent relays are typically characterized by their time–current curves, which define the tripping time as a function of the observed fault current. To maintain selectivity, the backup relay must operate after the primary relay by at least a predetermined coordination time interval \cite{purwar2017novel,saleh2017optimal}.

4) Recloser-fuse coordination: When a fuse is installed downstream of a recloser, the fuse’s minimum melting time must be several times greater than the fast tripping time of the recloser. This ensures that the recloser has the opportunity to clear the fault before the fuse operates, preventing unnecessary fuse replacements for transient faults \cite{10891366}.

5) Relay–fuse and relay–recloser coordination: When coordinating between a relay and a fuse, the relay should operate after the fuse’s maximum clearance time by at least a coordination interval. Similarly, when coordinating between a relay and a recloser, the relay should operate only after the recloser’s slow tripping mode has had sufficient time to clear the fault.

\section*{Methodology for DS Blackstart Restoration}

To solve the DS blackstart restoration problems, various methodologies can be applied, which can be classified into analytical approaches, data-driven approaches, and machine learning approaches.

\subsection*{Analytical Approaches}

Analytical approaches use optimization frameworks with objectives and constraints to determine solutions of DS blackstart restoration (Table \ref{tab:analytical_methods}).

\subsubsection*{Mixed-Integer Programming}

Mixed-integer programming (MIP) is a widely used optimization tool. In the context of DS blackstart restoration, discrete variables are employed to represent the binary states of electrical components, such as switches, circuit breakers, and tie-lines \cite{10798576}. Continuous variables are used to represent physical parameters, such as power flows, voltage magnitudes, and generation outputs. The combination of discrete and continuous variables makes the problem non-convex \cite{8606281}. For example, if activating a binary switch changes the network topology, the continuous power flows must satisfy a new set of constraints, which creates disjoint or piecewise feasible regions. For this reason, the MIP is highly sensitive to the problem size. To this end , advanced decomposition techniques, such as Benders decomposition \cite{9044401} and Lagrangian relaxation \cite{8605387} are often employed to enhance the problem scalability. MIP can be classified into mixed-integer linear programming (MILP) and mixed-integer nonlinear programming (MINLP). In MILP, both the constraints and the objective function are linear. The linear models are computational tractable, allowing the use of well-established optimization techniques such as branch-and-bound, branch-and-cut, and cutting-plane methods \cite{8362717,basu2023complexity,kobayashi2020branch}. In contrast, MINLP includes nonlinear constraints and objective functions. They are suitable to model nonlinear behaviours, such as AC power flows, inverter dynamics, CLPU behavior, and ESS operations. Solving MINLP problems often requires advanced algorithms, such as sequential quadratic programming, generalized reduced gradient method, and interior-point technique   \cite{frank2012optimal,9693496}.

\subsubsection*{Robust Optimization}

Robust optimization (RO) is an extension of MIP but can handle uncertainties such as unknown system damages, variability in renewable generation, and fluctuating demand. This is achieved by constructing uncertainty sets to represent these uncertain parameters \cite{yuan2016robust,xu2020dynamic,10538433}. The formulation adopts a min-max structure, which can be regarded as a stochastic game between two players. The ``min'' represents the decision-maker actions, such as minimizing load shed, while the ``max'' represents the actions of an adversarial opponent that selects the worst-case realization of uncertainty. Compared to traditional MILP, the introduction of uncertainty set makes the problem can not be solved by traditional branch-and-bound and branch-and-cut algoritms. To address this issue, cutting-plane method and column-and-constraint generation are developed \cite{zhang2022methodology}. These algorithms find the solution by iteratively adding constraints that are violated under the worst-case scenarios. However, RO can lead to conservatism since it is desinged for the worst-case scenario \cite{roos2020reducing}. A larger uncertainty set enhances robustness against a wider range of scenarios but often results in conservative solutions. For example, we may get a result to allocate more resources, such as additional generation reserves or backup ESSs. A smaller uncertainty set improves cost-efficiency but may compromise robustness in extreme conditions, such as sudden drops in renewable generation or spikes in load demand.

\subsubsection*{Stochastic Optimization}

There are two commonly used stochastic optimization: stochastic programming (SP) and distributionally robust optimization (DRO). SP optimizes the expected objective over a set of scenarios with probability distribution based on the Sample Average Approximation (SAA) method  \cite{10311549,7995099,9298836}. This can avoid the need to solve the problem over all possible realizations of uncertainty, which may be infinite and computationally infeasible. Also, as the number of sampled scenarios increases, the SAA solution is proven to converge to the true optimal solution. The major challenge of SP is the introduction of scenario-specific variables and constraints. To reduce the computational complexity, advanced algorithms, such as benders decomposition and progressive hedging algorithm (PHA) can be applied \cite{9576535}. In comparison, DRO formulates the problem in a robust way, allowing the probability distribution to vary within a predefined ambiguity set. Common techniques for defining this ambiguity set include Wasserstein distance \cite{9316222}, moment-based estimation \cite{yang2018distributionally}, and phi-divergence \cite{xie2020tractable}. By optimizing the worst-case expected objective over all possible distributions within the ambiguity set, the solution of DRO can be less conservative than RO, and more resilient than SP. To solve the DRO prblem efficiently, advanced solution techniques, such as column-and-constraint generation, cutting-plane methods, and dual reformulation, are often employed \cite{zheng2020data}.

\subsection*{Data-Driven Approaches}

The main difference of data-driven approaches from analytical approaches is their ability to leverage vast amounts of historical and real-time data to enable informed decision-making for DS blackstart restoration. This integration of real-world observations  bridges the gap between theoretical optimization frameworks and practical applications (Table \ref{tab:analytical_methods}).

\subsubsection*{Historical Data}

By analyzing past trends and extracting underlying patterns of renewable generation, failure analysis, and load consumptions, historical data can improve the performance of RO, SP, and DRO. For RO, historical data is used in constructing uncertainty sets that capture the range of worst-case scenarios observed in system behavior. For instance, the upper and lower bounds of DER outputs can be derived from historical data, which can be used to define the data-driven generation uncertainty set \cite{7352378}. For SP and DRO, historical data supports the estimation of probability distributions for uncertain parameters, used as empirical distributions to construct data-driven scenario sets for SP and data-driven ambiguity sets for DRO. For example, a scenario set for SP can be generated from outage data, where each scenario corresponds to a combination of line failures observed during past outages. Similarly, an ambiguity set for DRO can be constructed using historical load consumptions. The ambiguity set includes all probability distributions within a Wasserstein distance or satisfying moment-based constraints around the empirical distribution of historical load variations \cite{10113202,10709873}.

\subsubsection*{Real-time Data}

Real-time data can improve situational awareness during DS blackstart restoration. It supports dynamic corrective actions throught continuous monitoring of grid conditions \cite{6457435,ma2022robust}. For example, real-time data of renewable generation outputs, power flow measurements, voltage and frequency readings and load consumptions can serve as critical inputs to optimization frameworks. Accordingly, system operators can make data-driven restorative decisions that are more responsive to real-world scenarios. The integration of real-time data into decision-making contains several sequential steps. First, by utilizing advanced metering infrastructure, phasor measurement units, and intelligent electronic devices, the real-time data is collected and preprocessed. Second, the real-time data is fed into optimization frameworks, which are specifically developed to handle dynamic incoming data. Third, the optimization solutions, such as activating DERs or operating switches, are implemented in the system. Finally, real-time feedback on the results of these actions are collected again, using to refine solutions for the next iteration. To effectively manage sequential decision-making as real-time data becomes available, advanced tools and frameworks are required, such as rolling horizon optimization \cite{10787395}, model predictive control  \cite{8389207}, and markov decision process  \cite{10747032}.

\subsection*{Machine Learning Approaches}

Different from analytical approaches which depend on predefined mathematical models, or data-driven approaches that utilize data without incorporating a learning process, machine learning approaches analyze large and complex datasets to learn intricate patterns and make  decisions. The main machine learning approaches are supervised learning, unsupervised learning, and reinforcement learning.

\subsubsection*{Supervised Learning}

Supervised learning is the process of training models on labeled datasets, which clearly define the input-output relationships. Techniques such as decision trees, regression models, and neural networks are three types of supervised learning. For example, steady-state and transient data collected from distribution networks can be used as training inputs for decision trees. Then, the dynamic events, such as fault detection and load reconnection scenarios, can be classied \cite{al2018dynamic}. It helps system operators maintain situational awareness of the network state and execute corrective actions in real-time. In addition, historical outage associated with weather data can be leveraged to train regression models and neural networks to estimate average restoration times \cite{willems2024probabilistic,wang2023data}.The output can provide additional support for power dispatch and dynamic microgrid formation.

\subsubsection*{Unsupervised Learning}

Similar to supervised learning, unsupervised learning supports decision-making by analyzing collected data. However, the main difference is that unsupervised learning is used to identify patterns and structures in data without requiring labeled outputs. In this domain, techniques such as clustering, generative models, and dimensionality reduction are different types of unsupervised learning. For example, a generative model can be trained on historical renewable generation data, and then generate energy output scenarios for subsequent decision-making \cite{9862589}. Moreover, clustering methods can be applied to analyze historical outage data by examining time-series features, such as load disconnection and restoration rates \cite{jessen2022identification,wang2014time}. It helps decision makers to identify recurring behaviors and characterize the dynamics of outages. In addition, dimensionality reduction enhances decision-making by simplifying large datasets while preserving important information, such as those used in probabilistic power flow calculations.

\subsubsection*{Reinforcement Learning}

Compared to supervised and unsupervised learning, reinforcement learning (RL) can directly provide decisions for DS blackstart restoration. It allows agents to learn optimal policies through interaction with a stochastic environment. By employing trial-and-error exploration and exploitation, RL agents improve their policies to achieve blackstart objectives in an iterative way, such as minimizing restoration time or maximizing restored load. RL can be categorized into model-based and model-free approaches. A model-based RL requires an explicit model of the environment's dynamics to evaluate potential action trajectories before taking them in the actual environment. For example, RL agents can be trained on historical weather data to identify patterns associated with hurricanes, and explore various scenarios and develop policies for optimal dispatch of DERs during blackstart restoration \cite{hosseini2021resilient}. However, in real-world applications, deriving such an accurate model is challenging. To this end, Deep Q-Networks  \cite{huang2022resilient,igder2023service}, and Graph Reinforecement Learning  \cite{zhao2021learning,fan2024enhancing} which are model-free are developed. They provide a robust alternative by eliminating the need for explicit knowledge such as transition and reward functions.

\section*{Outlook}

The increasing integration of DERs driven by the renewable energy transition provides new opportunities to power systems worldwide. The feature that DERs can be deployed closer to the end users effectively reduces the restoration times and benefit both utilities and consumers. However, even though the main character of blackstart restoration has been transfered from large power plants to DERs, the current utilization of DERs still follows the traditional perspective of power system planning and operation. The potentials of DERs and associated advanced technologies in smart grid requires further exploration.

\subsection*{Advanced Dynamic Modeling and Protection}

The integration of DERs through inverter-based resources complicates the process of handling  transients. For example, abrupt load pickups and switching events can cause sharp voltage and frequency transients, while inrush currents from transformers and motor loads impose sudden stress on inverters. Despite advancements in inverter control techniques have enabled fast-response control to enhance local stability, integrating their transient behaviors into decision-making frameworks remains a major challenge. Future research can focus on dynamic models with explicit constraints or closed-form expressions that accurately represent inverter behavior within DS blackstart restoration problems. Furthermore, protection schemes needs further investigation. Conventional protection in DS depends on high fault currents and directional power flow for fault detection and isolation. However, in terms of inverters, short-circuit output is limited by control and hardware limits. This results in lower fault currents that challenge the operation of relays, fuses, and reclosers. In addition, the bidirectional power flow introduced by DERs, varying with generation and demand conditions, creates directional uncertainty, further complicating fault detection and protection coordination in DS blackstart restoration. To address these challenges, the development of reconfigurable protection mechanisms is required to dynamically adjust settings based on real-time system conditions. Also, fault characterization techniques to accurately capture the unique behaviors of inverter-based DERs including their controlled fault-current injection and fast response dynamics is necessary.

\subsection*{Situational Awareness and Real-Time Decision Support}

To deal with fast-changing status of DS blackstart restoration, situational awareness must integrate advanced sensing, predictive analytics, and intelligent decision-support systems. In this respect, distribution-level phasor measurement units, smart meters, and advanced sensors can provide high-resolution measurements of voltage, current, and frequency across the network. Also, real-time monitoring of abrupt load changes, inrush currents, and fluctuations in DER output can be revealed to system operators to support real-time decision-making. To this end, it is important to identify the optimal location and number for the installation of different types of sensing and measurement units, which is still an open topic. Future research can investigate the impact of sensor data from different locations on restoration and determine the optimal sensor upgrade scheme to improve DS blackstart performance. Another unresolved challenge is the intelligent decision support system that can handle vast, heterogeneous and asynchronous data from sensing and measurement units. These system should be capable of analyzing incoming data and update the restoration plan in an iterative manner. Moreover, finding optimal restoration solutions with computational time requirement during blackouts is difficult, especially for large-scale systems. Therefore, future research can be conducted on novel decision-making tools that combine classic optimization and machine learning approaches, such as reinforcement learning, deep adaptive dynamic programming, and online meta-learning. In addition, to further enhance the situational awareness, replicating non-convex constraints with tractable counterparts and predicting near-optimal solutions to warm-start optimization solvers based on real-time data needs further investigation.

\subsection*{Autonomous DS Blackstart Restoration}

As DERs become integral to the distribution level, they drive the evolution of the grid edge, which includes renewable generation, energy storage, and distributed computing operating closer to end users. In particular, advanced computing architectures such as edge computing and fog computing enables autonomous DS blackstart restoration, which can  overcome communication delays, computational bottlenecks, and the risk of single-point failures. However, to fully utilize these advantages, DERs, microgrid controllers, and local decision makers must independently assess system conditions and take coordinated actions. The blackstart procedure should proceed autonomously with the goal of maximizing the overall restoration performance while considering local restoration limitation. To achieve this antonomous, multi-agent reinforcement learning and federated learning offers a solution. Another challenge is that, local controllers may not intend to share sensitive data from security consideration, which means the system state can be partially observed. Therefore, secure and decentralized coordination mechanisms are necessary to preserve data integrity and privacy. In this respect, distributed ledger technologies such as blockchain act as a good start. Another advantage of autonomous DS blackstart restoration is scalibility. Most existing DS blackstart restoration strategies are designed for single-substation networks with only a few feeders. However, in terms of large service areas, such as a city-wide distribution system, there may be dozens of feeders, thousands of nodes, and hundreds of DERs. In such complex networks,  an autonomous and decentralized blackstart approach is necessary to improve scalability, computational efficiency, and response times.

\newpage

\begin{table}[H]\footnotesize
\caption{Grid-Forming and Grid-Following Inverters}
\label{tab:GFMI_GFLI_Comparison}
\begin{tabular*}{\textwidth}{@{\extracolsep{\fill}}p{4cm}p{6cm}p{6cm}}
    \hline
    & \multicolumn{1}{l}{\textbf{Grid-Forming Inverter}} 
    & \multicolumn{1}{l}{\textbf{Grid-Following Inverter}} \\
    \midrule
    Operating Principle & Voltage source, regulating voltage and frequency& Current source, synchronizing to an external  reference  \\
    \midrule
    Primary Energy Source & ESS (Common) / PV+ESS, Wind+ESS (Growing)
    & Solar PV, Wind (Common) / ESS (Sometimes) \\
    \midrule
    Blackstart Role & Initiating DS blackstart restoration\newline Establishing cranking patch to GFLIs& Supporting active and reactive power flow\newline Coordinating with GFMIs in load restoration \\
    \midrule
    Voltage Regulation & Active voltage control via $Q-V$ droop and VSG \cite{du2019comparative}  \newline Maintaining deviations within ±5–10\%  & Passive voltage regulation
    via reactive power support \newline  Operating in grid-supporting mode\\
    \midrule
    Frequency Stability & Active  frequency control via $P-f$ droop\newline Providing synthetic inertia via VSG (0.5s - 2.0s)\cite{cheema2020comprehensive}  & PLL synchronization-dependent\newline  Instability in weak grids or rapid frequency changes\\
    \midrule
    Fault Ride-Through \& Protection \cite{piya2018fault} & Enhanced ride-through with controlled inrush \newline Modest fault current support & Limited fault current support \newline May reduce output or disconnect under disturbances \\
    \midrule
    Key Advantages & Self-synchronizing, enhancing grid resilience, supporting weak networks, stabilizing renewables. &  Maximizing renewable energy harvest, low-cost deployment, and flexible grid integration. \\
    \midrule
    Limitations & Higher cost, complex control. & Unstable in islanded mode, cannot initiate blackstart \\
    \midrule
    
    Standards \& Compliance & IEEE 2800-2022, IEEE 1547.4-2011 & IEEE 1547-2018, UL 1741 SA \\
    \hline
\end{tabular*}
\begin{flushleft} ESS, energy storage system; PV, photovoltaic; GFMI, grid-forming inverter; GFLI, grid-following inverter; DS, distribution system; VSG, virtual synchronous generator; PLL, phase-locked loop. \end{flushleft}
\end{table}

\begin{table}[H]\footnotesize
\caption{Situational Awareness in DS Blackstart Restoration}
\label{tab:situational_awareness}
\begin{tabular*}{\textwidth}{@{\extracolsep{\fill}}p{2.5cm}p{7cm}p{7cm}}
    \hline
    & \multicolumn{1}{l}{\textbf{Renewable Energy Generation Forecasting}} 
    & \multicolumn{1}{l}{\textbf{Load Demand Prediction}} \\
    \midrule
    Roles/Purpose  
    & Guides optimal renewable energy dispatch, ESS operation, blackstart process, microgrid formation  
    & Supports load reconnection sequence, load shedding, cold load pickup mitigation  \\
    \midrule
    Key Approaches &&\\
    \midrule
    Stochastic Analysis  
    & Bayesian inference: Continuously updates probability distributions based on incoming meteorological data  \cite{ahmed2019review} \newline  
    Kalman filter: Processes noisy weather and sensor data to refine real-time generation forecasts  \newline
    Markov chain:  Models stochastic transitions and dependencies, effectively representing state-dependent variations
    & Regression analysis: Establishes  relationships between historical consumption patterns and external factors such as weather conditions, time of day, and seasonal demand variations \cite{li2014review} \\
    \midrule
    Machine learning
    & CNNs: Extracts spatial features from meteorological data \cite{heo2021multi} \newline  
    LSTMs / GRUs: Learns temporal dependencies
in time-series data, retains information over
long sequences \cite{xia2021stacked} \newline  
    GANs: Generates synthetic data that
resembles real-world scenarios, addresses data limitation  \cite{dong2022data}
    & Clustering algorithms: Identifies distinct load groups based on energy consumption behaviors \newline  
    Neural networks: Employs deep learning techniques to model nonlinear load behaviors based on diverse datasets \\
    \midrule
    Data Requirements  
    & High-resolution meteorological data, including solar irradiance, wind speed, temperature, and cloud cover \newline  
    Historical generation patterns, incorporating seasonal and diurnal variations in PV and wind power output \newline  
    Real-time sensor measurements, such as power output from DERs, turbine speed, and solar panel efficiency  
    & Historical load profiles, reflecting past consumption patterns under different environmental and system conditions \newline  
    Real-time measurements from advanced metering infrastructure and phasor measurement units, providing high-resolution data on voltage, current, and frequency 
    \\
    \hline
\end{tabular*}
\begin{flushleft}  CNNs, convolutional neural networks; LSTMs, long short-term memory networks; GRUs, gated recurrent units; GANs, generative adversarial networks; ESS, energy storage system; PV, photovoltaic; DERs, distributed energy resources. \end{flushleft}

\end{table}

\begin{table}[H]\footnotesize
\caption{Problem Formulation: Objectives, Goals, and Constraints in DS Blackstart Restoration}
\label{tab:problem_formulation}
\begin{tabular*}{\textwidth}{@{\extracolsep{\fill}}p{2cm}p{4cm}p{4cm}p{6cm}}
    \hline
    \multicolumn{1}{l}{\textbf{Problem}} & \multicolumn{1}{l}{\textbf{Description}} & \multicolumn{1}{l}{\textbf{Objective}} & \multicolumn{1}{l}{\textbf{Operational Challenges}} \\
    \midrule
    Power Dispatch \cite{liu2020collaborative} & Optimally allocate power from solar PV, wind, and ESSs to meet load restoration requirement while maintaining grid stability & Maximizes the total weighted restored load, minimizes restoration time, optimizes operational costs & Generation limits due to PV/wind availability\newline Inverter and ESS capacity constraints\newline SOC limits to prevent deep discharge or overcharging\newline Ramp rate limits to avoid sudden power fluctuations\newline Spinning reserve to ensure frequency stability \\
    \midrule
    Voltage Regulation \cite{10521743} & Maintain acceptable voltage levels while minimizing power losses and ensuring safe operation & Maintain voltage stability and reduce power losses & Voltage deviations must stay within acceptable range\newline Thermal constraints to prevent overheating\newline Inverter reactive power limits \\
    \midrule
    Demand Management \cite{wang2023sequential} & Control load restoration sequence, mitigate cold load pickup effects, and balance supply-demand to avoid overload conditions & Minimize supply-demand imbalance and prevent excessive inrush currents & Load restoration limits to prevent abrupt surges\newline CLPU constraints to limit transformer saturation and excessive current draw\newline Sequential restoration constraints to gradually re-energize loads \\
    \midrule
    Dynamic Microgrid Formation \cite{10858306} & Form self-sustaining microgrids that expand progressively while ensuring synchronization, stable voltage, and frequency control & Maximizes the restored load while maintaining frequency stability and minimizing the supply-demand imbalance at each step & Maintain radial feeder topology after reconfiguration\newline Synchronization constraints requiring voltage, frequency, and phase matching before reconnection\newline Constraints for controlled switch operations\newline Frequency stability limits to prevent RoCoF issues. \\
    \hline
\end{tabular*}
\begin{flushleft} CLPU, cold load pickup; RoCoF, rate of change of frequency; PV, photovoltaic; ESS, energy storage system; SOC, state of charge. \end{flushleft}
\end{table}

\begin{table}[H]\footnotesize
\caption{Optimization and Learning-Based Approaches for DS Blackstart Restoration}
\label{tab:analytical_methods}
\begin{tabular*}{\textwidth}{@{\extracolsep{\fill}}p{1.cm}p{4cm}p{6.5cm}p{4cm}}
    \hline
    \textbf{Category} & \textbf{Type} & \textbf{Application} & \textbf{Algorithm} \\
    \midrule
    {MIP} &  
    {Linear} &  
    Optimizes power dispatch, switching sequences under deterministic network conditions  
    &  
    Branch-and-bound, branch-and-cut, cutting-plane methods \\
    \cmidrule(lr){2-4}
    & {Nonlinear} &  
    Models nonlinear AC power flow, inverter dynamics, CLPU behavior, ESS operations  
    &  
    Sequential quadratic programming, generalized reduced gradient methods,  interior-point techniques \\
    \midrule
    {RO} &  
    {Interval Uncertainty Set \cite{liu2020bi}} &  
    Handles independent deviations in renewable generation  and load demand during restoration\newline Ensures robust decision-making by accounting for worst-case variations in known fixed bounds  
    &  
    Cutting-plane method, column-and-constraint generation \\
    \cmidrule(lr){2-4}
    & {Budgeted Uncertainty Set\cite{bertsimas2012adaptive}} &  
    Used when multiple uncertain parameters  are present, but only a limited subset is expected to deviate simultaneously\newline Essential for balancing conservatism and feasibility, unlike interval uncertainty which is too rigid  
    &  
    Benders decomposition, mixed-integer reformulation \\
    \cmidrule(lr){2-4}
    & {Data-Driven Uncertainty Set \cite{chen2020unified}} &  
    Defines uncertainty bounds using historical data\newline
    Enhances flexibility by reflecting actual system behavior instead of relying on predefined worst-case assumptions
    &  
    Depends on underlying structure \\
    \midrule
    {SP} &  
    {Finite Scenario Set} &  
    Handles predefined uncertainty realizations, such as load profiles and renewable generation, when a fixed set of scenarios sufficiently represents variability 
    &  
    Sample average approximation, progressive hedging \\
    \cmidrule(lr){2-4}
    & {Data-Driven Scenario Set \cite{wang2020data}} &  
    Utilizes historical trends and real-time data to update scenario probabilities, ensuring that uncertainty representations remain adaptive to evolving grid conditions
    &  
    Bayesian inference, deep generative modeling \\
    \midrule
    {DRO} &  
    {Moment-Based Ambiguity Set\cite{zhou2021distributionally}} &  
    Handles uncertainty in load demand and renewable generation by incorporating statistical moments (mean, variance, skewness) into the ambiguity set  
    &  
    Duality-based reformulation, conic programming \\
    \cmidrule(lr){2-4}
    & {Wasserstein Distance Ambiguity Set} &  
    Ensures robust restoration by considering worst-case deviations between empirical and actual probability distributions of uncertainties   &  
    Semi-definite programming, linear decision rules. \\
    \cmidrule(lr){2-4}
    & {Data-Driven Ambiguity Set \cite{10709873}} &  
    Historical data supports the estimation of probability distributions, forming empirical distributions \newline Real-time data supports the dynamic refinement of these ambiguity sets, enabling adaptive decision-making	    &  
    Depends on underlying structure  \\
    \hline
\end{tabular*}
\begin{flushleft} MIP, mixed-integer programming; RO, robust optimization; SP, stochastic programming; DRO, distributionally robust optimization; CLPU, cold load pickup; ESS, energy storage system; AC, alternating current.\end{flushleft}
\end{table}

\begin{figure}[H]
	\centering
	\includegraphics[width=7in,trim={0cm 11cm 0cm 0cm},clip]{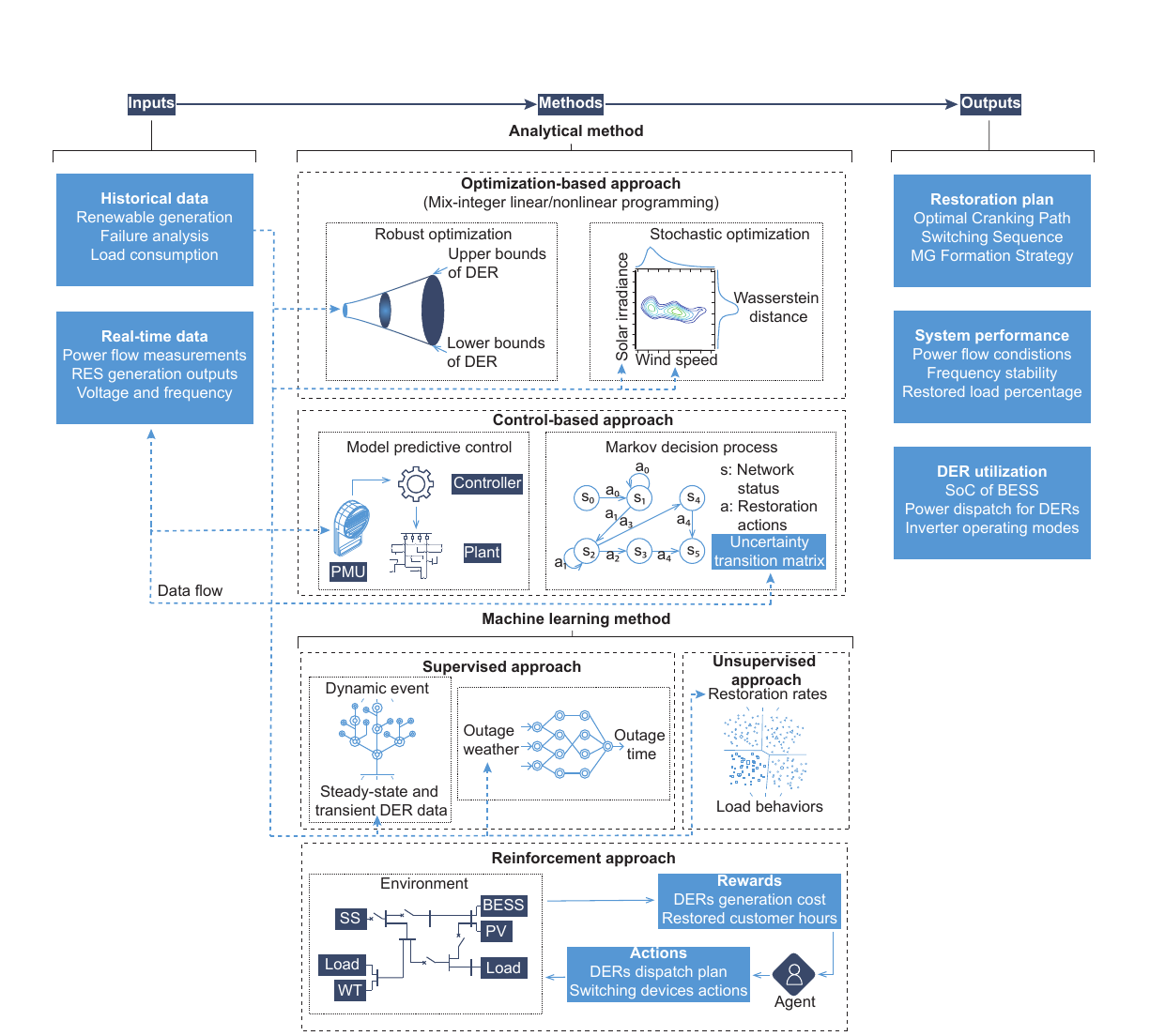}
	\caption{\textbf{Methodology for distribution system blackstart restoration.} DS blackstart restoration is addressed through analytical and learning-based approaches, both relying on historical and real-time data inputs. Analytical approaches formulate the problem into an optimization framework, using mixed-integer linear/nonlinear programming, stochastic optimization, and robust optimization to determine the optimal cranking path, switching sequences, and distributed energy resource (DER) dispatch under uncertainty. Control-based approaches, such as model predictive control, and Markov decision process dynamically adjust restoration decisions using real-time system feedback. Learning-based approaches leverage machine learning to enhance decision-making. Supervised learning predicts load recovery behaviors, while unsupervised learning clusters restoration scenarios. Reinforcement learning  models blackstart as a Markov decision process, where an agent learns optimal restoration sequences by maximizing predefined rewards. The outputs of these methodologies include the restoration plan, such as cranking path, microgrid formation, system performance metrics, such as restored load percentage, voltage and frequency stability, and DER utilization, such as energy storage system state of charge, and inverter operation modes.}
	\label{fig3}
\end{figure}

\newpage

\bibliography{reference}

\end{document}